\documentclass[preprint,5p,times,twocolumn]{elsarticle}

\usepackage{xcolor}
\usepackage{hyperref}
\usepackage{siunitx}
\usepackage{multirow}
\usepackage{verbatim} 
\usepackage{listings}
\usepackage{subfig} 

\usepackage{listings}

\usepackage{color} 

\definecolor{dkgreen}{rgb}{0,0.6,0}
\definecolor{gray}{rgb}{0.5,0.5,0.5}
\definecolor{mauve}{rgb}{0.58,0,0.82}

\usepackage{amssymb}
\usepackage{amsmath} 
\usepackage{mathrsfs}

\usepackage{pgf}

\journal{Nuclear Instruments and Methods A}
\begin{document}

\begin{frontmatter}



\title{\textbf{PyPWA: A Software Toolkit for Parameter Optimization and Amplitude Analysis}}


\author[inst1]{Mark Jones}
\affiliation[inst1]{organization={Norfolk State University},
            city={Norfolk},
            postcode={23504}, 
            state={VA},
            country={USA}}
\author[inst2]{Peter Hurck}
\affiliation[inst2]{organization={University of Glasgow},
            city={Glasgow},
            postcode={G12 8QQ},
            country={United Kingdom}}
\author[inst3,inst4]{William Phelps}
\affiliation[inst3]{organization={Christopher Newport University},
            city={Newport News},
            postcode={23606}, 
            state={VA},
            country={USA}}
\author[inst1,inst4]{Carlos W. Salgado}
\affiliation[inst4]{organization={The Thomas Jefferson National Accelerator Facility},
            city={Newport News},
            postcode={23606}, 
            state={VA},
            country={USA}}


\begin{abstract}
PyPWA is a toolkit designed to optimize parametric models describing data and generate simulated distributions according to a model. Its software has been written within the python ecosystem with the goal of performing Amplitude or Partial Wave Analysis (PWA) in nuclear and particle physics experiments.  We briefly describe the general features of amplitude analysis and the PyPWA software design and usage. We provide benchmarks of the scaling and an example of its application.
\end{abstract}



\end{frontmatter}

\tableofcontents 

\section{Introduction }

\subsection{A Brief Description of Amplitude Analysis}

In particle and nuclear physics, one of the main experimental goals is to determine the frequency with which a particular interaction  occurs. Experimental data are analyzed under the theoretical framework of quantum field theories. The data can be presented as counting histograms, in different bins of the particles' kinematic variables. The counting probability is represented by differential cross-sections that can be written, using Fermi's golden rule~\cite{Merz}, by

\begin{equation}  \frac{d\sigma}{d\tau} \propto  \sum_{ext.\;spins} \int |\mathscr{M}(\tau)|^{2} dx^{n}
\label{eqn:fermi}
 \end{equation} 
where $\mathscr{M}$, the transition amplitude, describes the physics of the particular interaction. The kinematic elements of particles as well as the Lorentz invariances are included in $dx^{n}$. The transition amplitude is a complex function that can not be measured directly; it has to be inferred from the cross-sections.
 In most cases, the amplitudes (therefore, cross-sections) depend on several kinematic variables, represented here by $\tau$~(e.g. beam energy,  the total mass of the final state set of particles, the angular distribution of the final particles, etc.). The experimentally measured kinematic variables are related to the intrinsic quantum numbers involved in the reaction. The main goal of the analysis is to extract those quantum numbers and, through them, gain information about the fundamental components of the reaction (i.e quarks and gluons). Of special importance are the angular momentum and spin quantum numbers, which are related to the angular distributions of the decay products. \par
 
 Different techniques are used to extract information about the properties of amplitudes by measuring cross-sections or quantities related to cross-sections. Dalitz analysis~\cite{Palano:2022ovo}  is used to study correlations between the kinematic variables involved in the reaction. These kinematic distributions are fit with a theoretically motivated parametric representation of the amplitudes. 
 In most amplitude studies, it is common to consider only the transition amplitude ($\mathscr{M}$) by binning the data in one or more of the kinematic variables $\tau'$, a subset of $\tau$, such that an {\it intensity}, $I$, is defined by
\begin{equation}  I(\tau') = \sum_{ext.\,spins}  |\mathscr{M}(\tau')|^{2}.  
\label{eqn:intenf}
\end{equation} 
The transition amplitudes $\mathscr{M}$ are defined by the model or theory.
When the intensity is only dependent on angular variables,  we can expand it in Harmonic Moments (Moments method)~\cite{CLAS:2008ycy}, or we can write the spin components using the Spin Density Matrix, the intensity is then parameterized by the matrix elements (SDME method)~\cite{GlueX:2021pcl}. Alternatively, we can expand the intensity in Partial Waves defined by the angular components (PWA method)~\cite{Salgado:2013dja,Ketzer:2019wmd}. These approaches are mathematically related and have their advantages and disadvantages. The PWA method directly obtains intensities classified by their angular quantum numbers and is more general but also more complex.

All of these methods, in practice, are reduced to the optimization of amplitude model parameters to describe the data. Therefore, any software infrastructure used for these types of analyses has to perform the parameter optimization (from now on referred to as fitting) and the generation of modeled or simulated data.

 The PyPWA software aims to fill those necessities by providing the user with a flexible and modular software toolkit to define any type of amplitude and to represent the data by any set of variables. Several examples and tutorials are included with the software and documentation. One example of PWA using the PyPWA package for the reaction $\gamma p \rightarrow pX \rightarrow p\eta\pi$ is presented in~\ref{sec:example}.

\section{Toolkit requirements} \label{sec:SimFitPred}
\subsection{Simulation, Fitting, and Prediction}
The main tasks the toolkit performs are the generation of simulated events, the optimization of parametric models, and the generation of simulated data based on the optimized model (from here on referred to as prediction).

The generation of simulated data by PyPWA is performed by two methods. Data are simulated through the rejection sampling method from a model with a priori parameters (production mode). Production simulation can be performed to improve the understanding of the analysis or to design an experiment. In production mode,  the software can be used to simulate any combination of resonances and waves. Alternatively, the simulation is used to predict the properties of the data according to the optimized model allowing the predicted kinematic properties to be compared to the data. Additionally, this can be folded in with detector acceptance incorporated through an independent simulation, e.g. Geant4~\cite{Geant}.

One way to optimize the model, implemented in the software, is the maximization of the extended likelihood that is defined as~\cite{Salgado:2013dja}
\begin{equation}
\mathscr{L} = \Big(\frac{\mathscr{N}^{N}}{N!}e^{-\mathscr{N}}\Big) \prod^{N}_{i}I_{i}({\tau}_{i}',\overrightarrow{V})
\end{equation}
where $N$ is the number of data events and $\overrightarrow{V}$ is the set of model parameters. The total number of expected events, $\mathscr{N}$, calculated using simulated events, is defined as 
\begin{equation}  \mathscr{N }= \eta \frac{1}{N_{a}} \sum^{N_{a}}_{j} I_{j}({\tau}_{j}',\overrightarrow{V})\end{equation} 
where $\eta$ is the detector acceptance and $N_{a}$ is the number of accepted simulated events.
For numerical reasons, and because most available optimizers minimize a loss function, it is more efficient to use the negative logarithm of the likelihood. To obtain the optimal $\overrightarrow{V}$ values, the following is minimized

\begin{equation}  -ln\mathscr{L} = -\sum_{i=1}^{N}  ln I_{i}({\tau}_{i}',\overrightarrow{V}) +  \mathscr{N }.
\label{eqn:like}
\end{equation}

Optimization can be done with a variety of packages available within PyPWA. We provide an interface to iMinuit~\cite{iminuit,James:1975dr}, and emcee\cite{Foreman_Mackey_2013} for cases where a Markov Chain Monte Carlo (MCMC) method is preferred (see Sec.~\ref{sec:optimizer}).

\subsection{Other packages}
Prior to the development of PyPWA other photoproduction PWA packages existed such as PWA2000~\cite{Cummings2003AnOAOriginal}. These packages were predominantly written in C/C++ and inspired PyPWA to choose to develop in Python for the increased usability. Some other examples that were developed both before and while PyPWA was developed include ComPWA~\cite{compwa}, AmpTools~\cite{amptools}, GPUPWA~\cite{gpupwa}, and ROOTPWA~\cite{rootpwa}. The ComPWA project started with using C++ and recently developed into a Python based toolkit that provides a modular and flexible package for PWA. AmpTools is a C++ toolkit that focuses on automation and uses CUDA~\cite{Nickolls:2008gqs} for GPU access.

\section{Software Design and Implementation}
PyPWA was designed to be a flexible set of tools within the Python ecosystem to fit multi-dimensional models and generate simulations. It uses an object-oriented design for data structures and components that must store a runtime state or support a plugin-like design, with a functional design for the remaining package. 

PyPWA is a package built from individual and mostly independent components, which yields the flexibility of a toolkit design, allowing for use beyond its original scope. PyPWA has two main components: data processing and data analysis. For data processing, PyPWA contains its own set of libraries for parsing, masking, and operating directly on data for multiple different data types and underlying libraries. For data analysis, there are tools to aid in developing likelihoods, amplitudes, fitting, and visualization of the data.

\subsection{Choice of Language}
PyPWA aims to work with existing workflows (e.g., CERN's ROOT~\cite{root}) without forcing its design principles onto the user, allowing for the rapid development of new amplitudes for analysis. Python was selected because it already had the necessary tools and libraries in its ecosystem dedicated to numerical processing and analysis, such as NumPy~\cite{harris2020array}, SciPy~\cite{2020SciPy-NMeth}, and Matplotlib~\cite{Hunter:2007}.  In addition, Python provides support for binding to other languages with tools such as F2PY~\cite{peterson2009f2py} for Fortran or Cython~\cite{behnel2011cython} for C/C++. Finally, the nature of Python's interpreted language allows for high-level development without the complexities of resource management, as well as tools that allow for iterative block-level development through Jupyter~\cite{soton403913} or iPython~\cite{PER-GRA:2007}.

One limitation of Python is that it requires a Global Interpreter Lock (GIL), a mutex in the CPython implementation of Python that synchronizes access to Python objects, preventing multiple native threads from executing Python bytecodes concurrently. To bypass the GIL, the Multiprocessing module uses forks for the parallel execution of Python code. Multiprocessing is a key feature of PyPWA discussed in Sec.~\ref{sec:ppm}. Python can take advantage of libraries compiled in lower-level languages, e.g. C and Fortran, which are capable of low-level optimization such as using vector instructions. NumPy, PyTorch~\cite{NEURIPS2019_9015}, Matplotlib, and iMinuit are examples of the optimized libraries that are used.  NumPy and SciPy allow for the development of amplitudes that PyPWA can automatically scale to all available hardware threads. Packages like Cython and F2PY allow hybrid amplitudes that leverage Python, C, and Fortran components.

In order to increase the speed of the fits, there are three ways to boost performance: Multithreading, vectorized instructions, and general-purpose computing on GPUs (GPGPU). Modern CPUs have more than one core, with eight cores being common, and this is the first place in which performance can be increased by using all available cores with multiprocessing. Some architectures, such as the x86 architecture, have vector instructions as a part of the CPU instruction set, also known as ``Single Instruction/Multiple Data" (SIMD), which allows the processor to perform multiple floating point operations in a single clock cycle. NumPy has built-in support for vector instructions allowing the user to write vectorized code in Python seamlessly. The final way to increase the performance of PyPWA is to offload the computations to GPGPU. For GPGPU calculations, the PyTorch Tensor Library supports AMD and NVIDIA GPGPUs, as well as Apple's Silicon, allowing for a broad range of supported hardware. Without these libraries, the user would be required to write their code in C/C++ or Fortran to achieve similar performance levels.

Finally, as an interpreted language, Python allows for interactive development similar to tools in ROOT. Two tools support interactive development: iPython and Jupyter notebooks. The iPython interpreter is similar to ROOT's CLING C/C++ interpreter and is executed from the command line. Jupyter notebooks are web browser based and have rearrangeable code cells that can be executed in any order and provide markdown blocks for inline documentation. Matplotlib plots and standard output of the python code are displayed below the code cell in which they are written.

\subsection{Data Preparation}
\begin{figure}
    \centering
    \includegraphics[width=\linewidth]{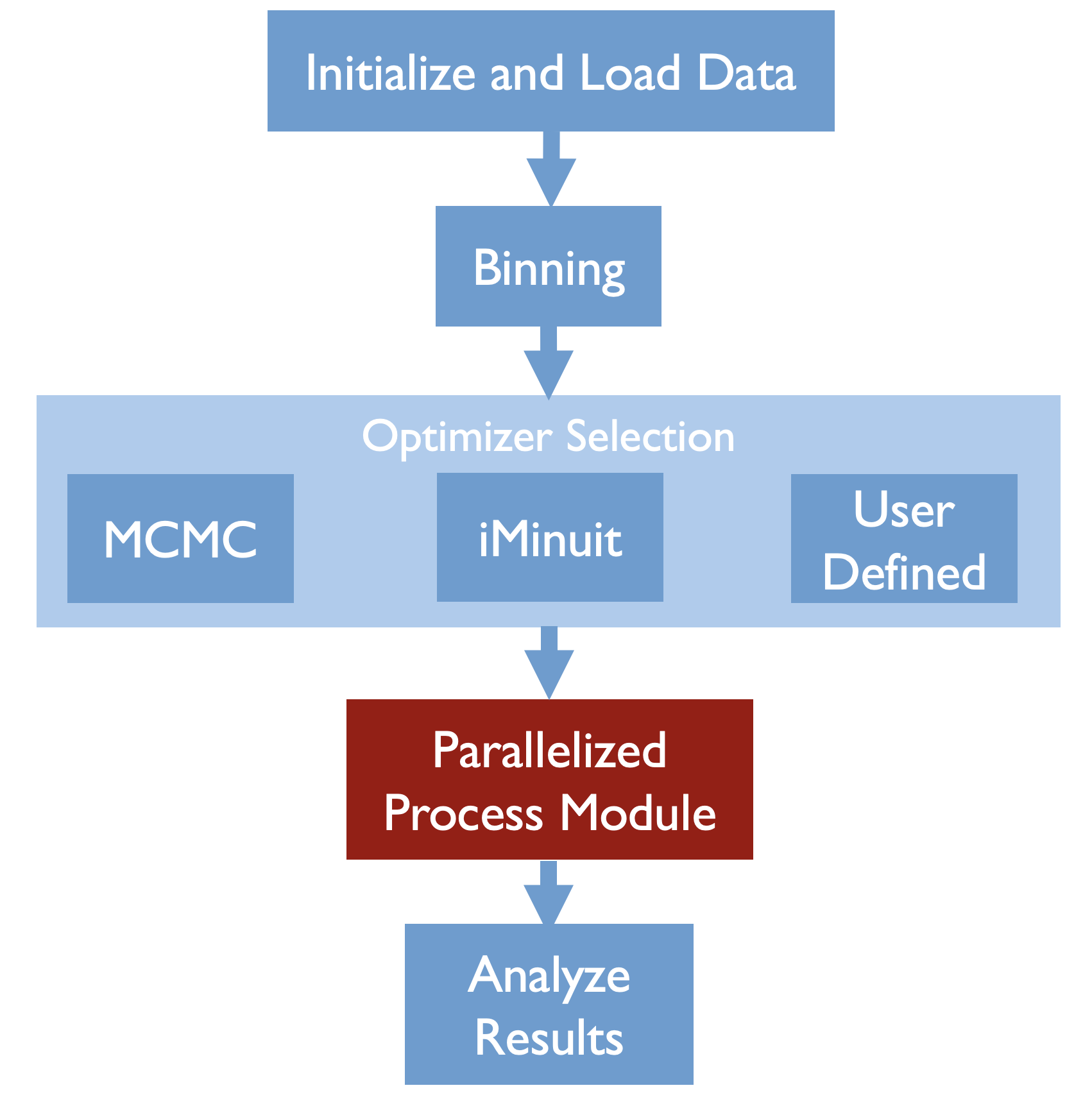}
    \caption{Process flow control diagram for PyPWA shows the steps from loading to analyzing the data. The parallelized process module is described in Figure~\ref{fig:pypwa_parallelism}.}
    \label{fig:pypwa_flowchart}
\end{figure}

Figure \ref{fig:pypwa_flowchart} describes the entire fitting process with PyPWA. After initial event selection, data can be loaded in the following formats: GAMP~\cite{Cummings2003AnOA}, CSV, NumPy, or user-defined format representing a 2-dimensional array or a tree of 4-vectors. The data module leverages the plugin architecture for built-in and user-defined formats to implement the data caching described below.

While PyPWA data module is the preferred most compatible way to handle data, it is not the only way. Pandas and NumPy can load data to operate with the toolkits since the toolkits supports Panda's DataFrame, NumPy's structured arrays, and PyTorch's Tensors as data types for operation.

The default loading and writing mechanisms have additional features not found in the other tools, specifically plugins and caching. The data module inside PyPWA supports plugins that can define readers and writers, parsers, and dumpers. A reader and writer will read and write a single event to and from a file, whereas a parser and dumper will read the entire file into a structured array or write an entire array out to a file. By supporting plugins, we can allow other developers to define file formats that best suit their needs and adapt PyPWA's parser to their file format while leveraging the rest of the toolkit. Lastly, any file written or read with PyPWA's data module is cached using Python's pickle module. Inside the cache is a SHA512~\cite{cryptoeprint:2010/548} sum of the original file, and if the sum does not match the file's current sum, the cache is invalidated, parsing the source file again.

The data module will return all parsed data in either NumPy structured arrays or Panda's DataFrames. DataFrames are similar to structured arrays but have additional methods that allow direct data visualization and syntax for filtering and searching. NumPy structured arrays focus on numerical computing performance and, as such, have no tools for data visualization. We provide utilities to swap between these two data structures to allow users the best of both of these tools.

Finally, many users may need to split datasets into subsets, called bins, in one or more variables, e.g. bins of energy or mass. PyPWA supports multiple binning functions by default, allowing bins to be split by a given range or by a fixed number of events. For both of these functions, the user provides a dataset to be binned, and the function will return a list of bins. If binning by range, the bins returned will all contain the same widths in the data, with the remaining data included in the first and last bins, decided using a single specified variable from that dataset. When binning by a fixed number of events, each bin will have the same number of events, again with the remaining data included in the first and last bins.

\subsection{Parallel Processing Module} \label{sec:ppm}
The Global Interpreter Lock (GIL) mechanism hinders Python's parallel processing ability by restricting the parallel execution of python on one interpreter. As a way to bypass the GIL, multiprocessing is used to enable multiple interpreters to run concurrently. However, utilizing multiprocessing, which uses forks, presents technical challenges due to the fact that each process operates independently without shared memory, requiring inter-process communication through Operating System pipes. Specifically, two pipes per process are necessary to enable bi-directional communication.

\begin{figure}
    \centering
    \includegraphics[width=0.45\textwidth]{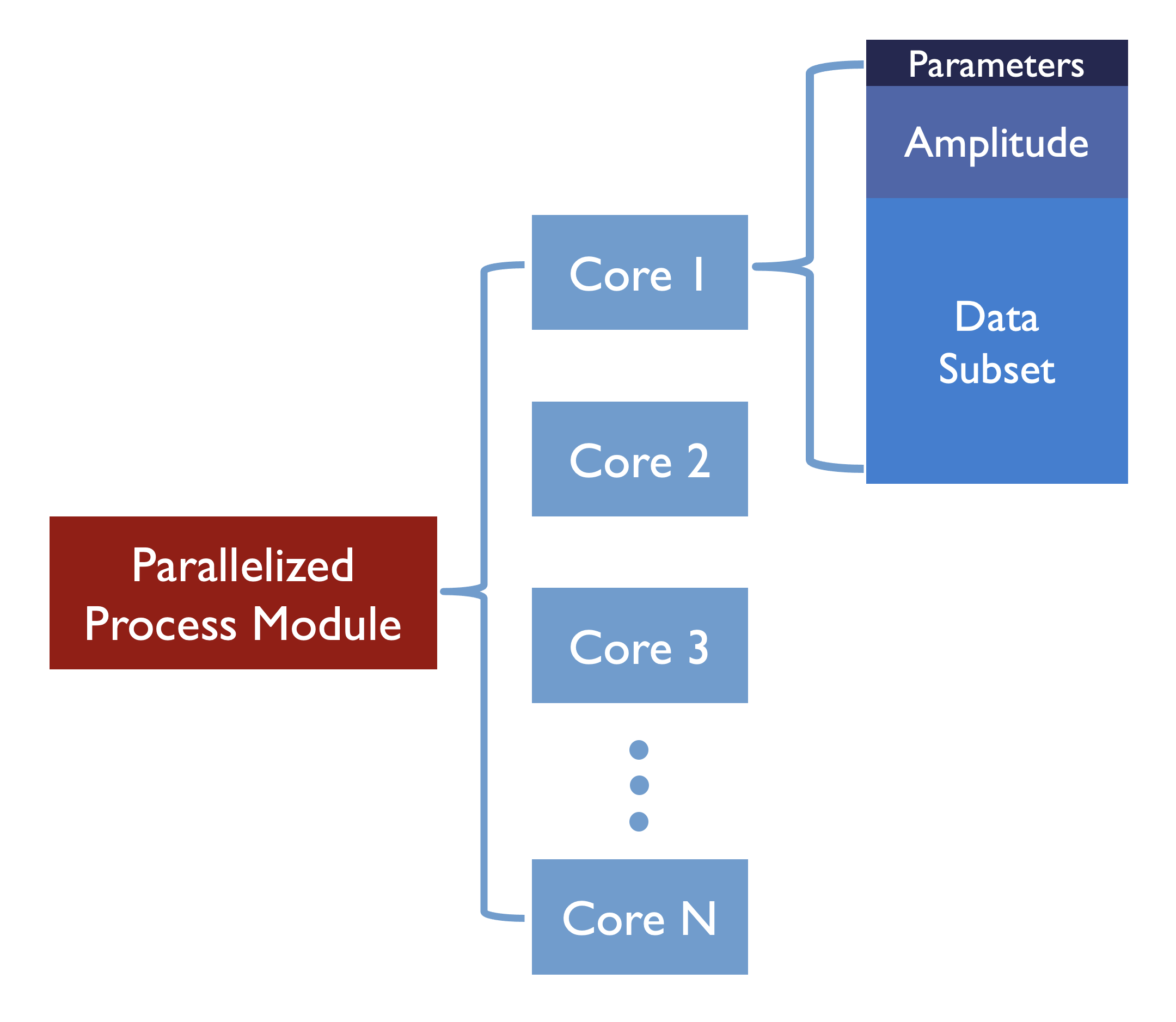}
    \caption{The Parallel Processing Module in PyPWA leverages multiprocessing to perform CPU-bound tasks concurrently, enabling scaling across all available hardware threads while bypassing the limitations of the GIL. Multi-GPU fitting uses multithreading instead of multiprocessing and one core in this diagram would correspond to one GPU.}
    \label{fig:pypwa_parallelism}
\end{figure}

PyPWA implements multiprocessing by inheriting from the Process class in the Multiprocessing module. The features that are added include duplex/simplex communication and kernels. In PyPWA, a kernel refers to a unit of execution that contains user-defined amplitude and its assigned data subset. The processing module is a processing factory that takes a kernel and dataset and then scales those across a number of processes, defaulting to the total number of hardware threads on the host system. The processing factory duplicates the provided kernel for each number, N, of selected hardware threads, splits the dataset into N sub-datasets, and then attaches each sub-dataset into one duplicated kernel as shown in Fig. \ref{fig:pypwa_parallelism}. The kernels, with their data payloads, are then assigned to a child process. After the processes are forked, the processing factory returns a communication object that passes values to all processes using pipes and processes the resulting values. The user primarily interacts with the processing module in the form of the likelihood module and simulation module.

Multiprocessing with forks works for many computational workloads, however, some libraries such as PyTorch and CUDA are incompatible with forks. In order to provide multi-GPU support, the kernels use threads and similar scaling to the CPU scaling is seen with multiple GPUs as discussed in Section \ref{sec:benchmarks}.

The processing module combined with the instruction-level vectorization in NumPy and PyTorch allows for high scalability across hardware resources. This scalability is built directly into PyPWA and does not require any additional input from the user.

\subsection{Optimizer Implementation}\label{sec:optimizer}
As introduced in Section~\ref{sec:SimFitPred} a loss function, such as a negative log-likelihood function, is minimized to extract model parameters which best fit the data. PyPWA requires the user to provide the intensity function $I(\tau',\overrightarrow{V})$ (\textit{c.f.} Eq.~\eqref{eqn:intenf}) which is used when fitting the data. This should be done in the form of a python class that contains at least two methods: \verb+setup+  and \verb+calculate+. The first will handle the data input and initialization, while the second performs the actual calculation and returns the intensity as a function of the parameters. PyPWA will use this python class in the likelihood object it passes to an optimizer. \par
In general, an optimizer is required to be able to accept the loss function provided by PyPWA as input together with the set of parameters. It should then be able to return the final set of parameters that optimize the loss function together with their associated uncertainties. Ideally, one would like to be able to also constrain the range of individual parameters in the optimization. It should be possible to use any optimizer that has this capability within PyPWA. \par
PyPWA provides two built-in optimizers which are described in the following. Due to its modular nature it is possible to add additional optimizers.

\subsubsection{iMinuit}\label{sec:iminuit}
A common tool to perform fits to data in high-energy physics is MINUIT2 (based on~\cite{James:1975dr}). PyPWA utilizes the Python implementation iMinuit~\cite{iminuit}. \par
MINUIT2 itself is not a minimizer but provides a range of algorithms that perform the minimization. The most common choice is MIGRAD, which uses the first and second derivatives of the function to be minimized in order to find the best set of parameters and approximate uncertainties. In a follow-up step the user can use the HESSE or MINOS algorithms to improve the estimation of parameter uncertainties. For more information on the exact working and best use cases for each algorithm, consult Ref.~\cite{James:1975dr} or the MINUIT2 User's Guide~\cite{James:2004xla}.\par
PyPWA provides an interface which takes the user-defined intensity function, i.e. the Python class, and input data and automatically turns it into a loss function which is passed to iMinuit. The following code snippet demonstrates this using a negative log-likelihood:

\begin{lstlisting}
with pypwa.LogLikelihood(Func(),data) as nll:
    minimizer = pypwa.minuit(
       fit_settings,
       nll
    )
\end{lstlisting}

The \verb+minimizer+ object returned by \verb+pypwa.minuit+ is iMinuit's \verb+Minuit+ object. That means the user now only needs to call the minimizer method of choice, e.g. \verb+minimizer.migrad()+, to perform the fit.\par
iMinuit provides some useful utility methods to inspect the final results. It is recommended that the user should carefully inspect the set of best parameters and uncertainties provided by iMinuit. It should be noted that HESSE and MINOS packages provide a better uncertainty estimation than MIGRAD and careful study of the uncertainties may be required.

\subsubsection{\textit{emcee} - Markov Chain Monte Carlo}\label{sec:emcee}
Instead of using a minimizer, PyPWA offers the user the option to perform parameter estimation via Markov Chain Monte Carlo (MCMC). For this, the Python package \textit{emcee}~\cite{Foreman_Mackey_2013} is used.\par
In contrast to a minimization of a loss function, MCMC explores the possible parameter space. A range of different algorithms to do this is provided by \textit{emcee}. In general, they can be split into the following stages:
\begin{enumerate}
    \item Choose a set of starting parameters $\vec a_0$
    \item propose a new set of parameters $\vec a_{n+1}$ (step)
    \item calculate likelihood
    \item accept or reject the proposed step based on likelihood\\
        if accepted: add $\vec a_{n+1}$ to output chain \\
        if rejected: add $\vec a_{n}$ to output chain
    \item go back to 2.
\end{enumerate}
Repeat this for a certain number of steps $n$.\par
The proposal of a new set of parameters $\vec a_{n+1}$ is often done at random based on the previous step $\vec a_{n}$. One possibility is to choose $\vec a_{n+1} = \vec a_{n} + \mathcal{N}(0,\vec\sigma)$, where each parameter gets modified by a random number drawn from a normal distribution $\mathcal{N}$. The set of widths $\vec\sigma$ which determine how much each parameter is changed at every step is called step size. Whether a new step is accepted depends on the likelihood of the current ($n$) and new step ($n+1$). It is not advisable to always accept steps that increase the likelihood as this might cause the algorithm to get stuck in local maxima.  In order to avoid local maxima, steps that decrease the likelihood should be accepted with a certain probability.\par

\begin{figure}[h]
    \centering
    \includegraphics[width=\linewidth]{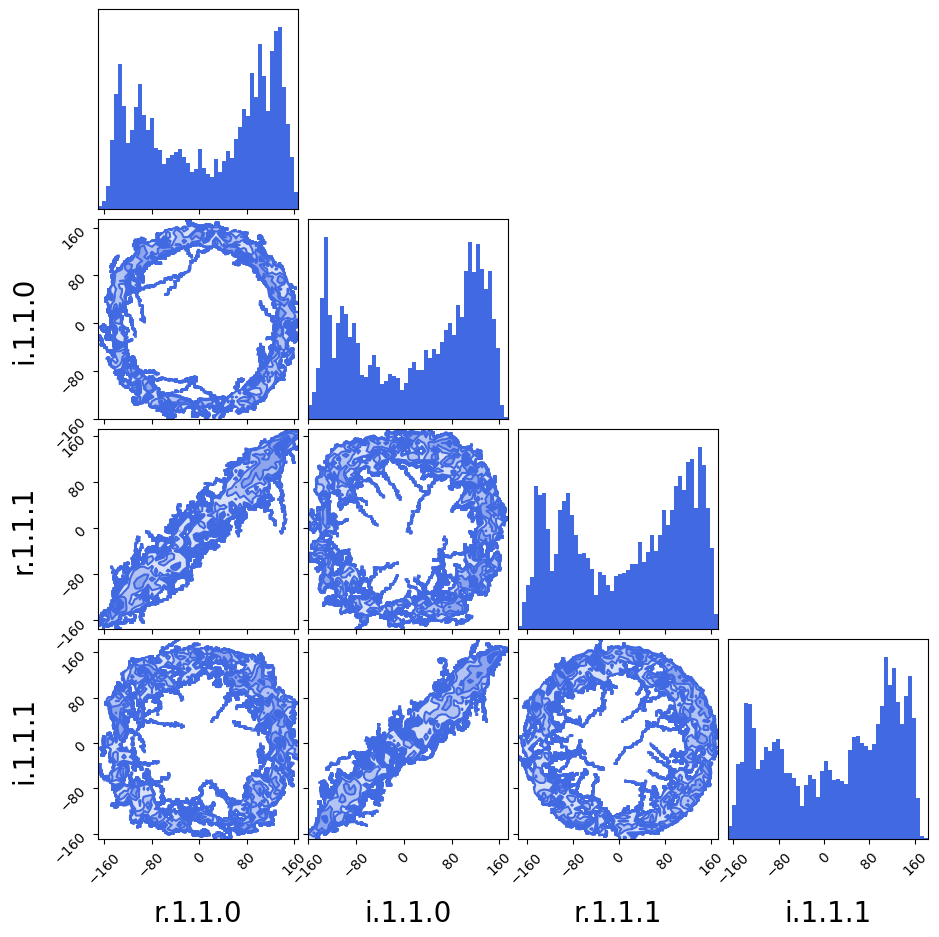}
    \caption{Corner plot for a sub-sample of the parameters used in the example shown in~\ref{sec:example}. The axes show the real (r) and imaginary (i) part of the production amplitudes abbreviated by their quantum numbers $\epsilon$, $L$, and $m$ in arbitrary units. The plot nicely visualizes correlations between the parameters. It was produced using the {\it corner} package~\cite{corner}.}
    \label{fig:corner_plot}
\end{figure}

PyPWA provides the interface to access \textit{emcee's} MCMC sampler. Several different algorithms (moves) which propose and accept steps are included. The main strength of \textit{emcee} is in so-called \textit{ensemble sampling} in which multiple chains are run in parallel.  For the full documentation of all moves, Ref.~\cite{Foreman_Mackey_2013} should be consulted. PyPWA returns the \verb+EnsembleSampler+ object which can be used to retrieve all the available information from \textit{emcee}. Most importantly, it can be used to get the resulting Markov chain, the complete history of all steps. This chain can be examined to study correlations between parameters and to extract a set of parameters that best describe the data together with their uncertainties. A useful tool to investigate the resulting chain is the so-called \textit{corner plot} which can be created using the \textit{corner} package~\cite{corner}. An example for such a corner plot is shown in Figure~\ref{fig:corner_plot}.
It shows a sub-sample of the parameters used in the example in~\ref{sec:example}. The parameter values for all individual steps in the chain are plotted against each other and reveal correlations. The corner plot also produces the one-dimensional projections which are often useful.\par
When using MCMC to perform parameter estimation it is important to study the generated chains carefully. It is necessary to ensure that the chains run long enough to converge. Although a step only depends on its immediate predecessor, there will be some auto-correlation within the chains, because steps are usually small enough to keep the acceptance rate reasonably high. This auto-correlation needs to be accounted for when information is to be extracted from the chain based on statistically independent samples. A common way to achieve this is to \textit{thin out} the chain. That means that, depending on the auto-correlation length, e.g., only one in every 10, 20, or 50 steps of the chain is used for parameter estimation.

\subsection{Documentation}
In PyPWA, the documentation is contained within the source code allowing Python's docstrings to grow and mature with the code leveraging PEP 256~\cite{pep256} and PEP 287~\cite{pep287}, which define Python's included support for literate programming. The docstring format used in PyPWA was initially defined for NumPy and blends an easily parsed markdown format for documentation generators while remaining syntactically simple to aid reading prior to rendering.

Python's built-in help function lets users view the docstrings attached to modules, classes, and functions directly from an IDE, iPython or Jupyter, which provides an interactive, built-in manual to PyPWA. For users and developers there is another tool inside Python's ecosystem, \textit{Sphinx}, that can parse the docstrings and render the documentation into a static website, which \textit{Read the Docs} currently hosts for PyPWA.

Documentation can be found on the PyPWA \textit{Read the Docs} ~\cite{docs} and the website ~\cite{website}. The full source code can be found on the GitHub page~\cite{github}.
\section{Benchmarks} \label{sec:benchmarks}
The benchmarks presented in this work evaluate the scaling of PyPWA fits on multi-core CPUs and multiple GPUs. The benchmarks were performed on a dual-socket Intel Cascade Lake system with three NVIDIA V100 GPUs. A detailed overview of the system specifications can be found in~\ref{sec:specs}. The benchmark only considers the calculation of the loss function as it is typically the most computationally expensive part of the fit. The data loading and visualization tasks are one-time tasks that typically have a negligible effect on the runtime. The optimizer can have a variable number of function calls to the loss function, which necessitates the exclusion of these tasks to provide uniform benchmark conditions. 

The benchmark procedure involves recording the execution time for a loss function call based on \ref{sec:example}, which is intended to mimic a fit, that uses a predetermined sequence of simulated fit parameters that are cached for the selected benchmark amplitude. This synthetic benchmark has 11 million events for both simulated experimental data and accepted data in the loss function and the number of loss function calls were set to 2636 calls, which had previously been observed in a fit performed with the default optimizer. This approach ensures that the benchmark has uniform conditions for all runs. The only parameter that varies is the number of threads utilized for the loss function calculation. Nonetheless, there exist uncontrolled factors in the benchmark testing process due to the variability in the server environment and variable CPU clock frequency which is known as Intel Turbo Boost for Intel processors that implement this feature.

\begin{figure}
    \centering
\includegraphics[width=\linewidth,trim={0.0cm 0.0cm 0.3cm 0.3cm},clip]{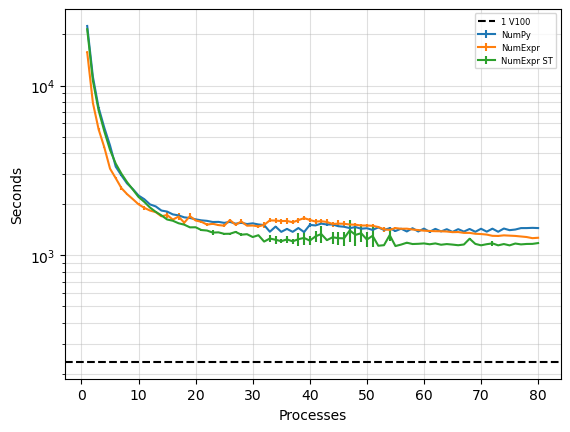}
\caption{The total execution time, using different libraries,  is shown as a function of the number of processes. For additional comparison, a single V100 GPU data point is shown as a horizontal black dashed line. The benchmark fit was based on \ref{sec:example}.}
    \label{fig:execute_time}
\end{figure}

The benchmark system had two Intel CPUs with forty threads each and three NVIDIA V100 GPUs and is described in Table~\ref{tab:specs}. In order to measure the scaling there was one benchmark configuration for each thread available on the system as well as one configuration for each GPU. Each configuration was tested four times and the mean and standard deviation of the benchmark execution times were recorded. Figure~\ref{fig:execute_time} displays the execution time for each CPU benchmark configuration and a single NVIDIA V100 for comparison.

We determine the theoretical speedup of the processing module using some of the most common mathematical libraries in Python: NumExpr, NumPy, and PyTorch for GPU processing. NumExpr has its own internal scaling mechanism which provides an analog to the PyPWA processing module; for this reason, NumExpr is included twice. NumExpr is included with its own multithreading enabled or disabled (NumExpr ST), both using the Parallel Processing Module. NumExpr ST and multiprocessing with PyPWA provide the best performance overall on CPUs as NumExpr is able to evaluate expressions without returning intermediate results and PyPWA supports data caching. However, using a single V100 GPU performs much better than the fastest 80 thread CPU execution.

The benchmark data is fitted with the functional form of Amdahl's Law~\cite{10.1145/1465482.1465560}. The theoretical speedup is given by
\begin{equation}
S_{latency}(s) = \frac{1}{(1-p) + (\frac{p}{s})}
\label{eq:amdahl}
\end{equation}
depending on $s$ the number of threads. The parameter $p$ in Amdahl's law (Eq.~\eqref{eq:amdahl}) is the proportion of the algorithm that is parallel. The closer it is to 1, the better the algorithm will scale. The parameter $p$ has been determined using the benchmarks. 

Table~\ref{tab:benchmark_results} displays the resulting values for speedup ($S_\text{max}$) and percentage of the algorithm that is parallel ($p$) for each math library for 80 threads on the two CPUs or the three GPUs. Table~\ref{tab:benchmark_comparison} shows a direct comparison between single thread execution and execution on the whole CPU using all 80 available threads. With an almost 96\% parallel algorithm on CPUs and almost 99\% on GPUs, these results show that PyPWA scales well on the benchmark system.

\begin{table}[]
    \caption{Results for the benchmarks as described in the text. The execution time is specified for 80 threads (or three GPUs in the case of the V100). Also listed are the speedups $S_\text{max}$ as defined in the text and ratio $p$ as defined in Eq.~\eqref{eq:amdahl}.}
    \small
    \centering
    \begin{tabular}{cccc}
        \hline
        \multirow{2}{*}{Math Library} & Execution Time & \multirow{2}{*}{$S_\text{max}$} & \multirow{2}{*}{$p$} \\ 
         & (in seconds) &  &  \\ 
        \hline

        NumPy & $1447.25 \pm 2.48$ & $15.51 \pm 0.03$ & $94.74\%$ \\
        NumExpr & $1268.83 \pm 2.14$ & $12.40 \pm 0.02$ & $93.10\%$  \\
        NumExpr ST & $1180.56 \pm 2.02$ & $18.19 \pm 0.03$ & $95.70\%$  \\
        PyTorch: 3 V100s & $80.70 \pm 2.16$ & $2.92 \pm 0.08$ & $98.61\%$ \\
        \hline
    \end{tabular}
    \label{tab:benchmark_results}
\end{table}

\begin{table}[]
    \caption{Comparison of execution times for a single thread and 80 threads using various math libraries described in the text. In the case of the V100 only one CPU thread was used.}
    \small
    \centering
    \begin{tabular}{ccc}
        \hline
         & Execution Time & Execution Time \\ 
        Math Library & one process & 80 processes  \\ 
         & (in seconds) & (in seconds)  \\ 
        \hline
        NumPy & $22440.85 \pm 13.13$ & $1447.25 \pm 2.48$  \\
        NumExpr & $15731.39 \pm 115.31$ & $1268.83 \pm 2.14$   \\
        NumExpr ST & $21473.77 \pm 46.79$ & $1180.56 \pm 2.02$   \\
        PyTorch: 1 V100 & $235.53 \pm 5.81$ & - \\
        PyTorch: 3 V100s & $80.70 \pm 2.16$ & - \\
        \hline
    \end{tabular}
    \label{tab:benchmark_comparison}
\end{table}

\section{Summary}
 PyPWA provides a flexible set of tools within the Python ecosystem for amplitude analysis of multi-particle final states. The package is built from individual and mostly independent components that the user can arrange in a variety of ways. PyPWA primarily functions as a toolkit and can solve a broad collection of optimization problems. Users provide a function (model), data, and simulation in their preferred data formats. PyPWA provides tools for data processing and analysis. It also provides various ways of speeding up the calculations through parallel processing and user-friendly GPU support with PyTorch.\par
 
 The flexibility of PyPWA and the use of many standard Python packages make it an ideal tool to perform fits of models to data. The packages are user-friendly to install on Linux and MacOS using Anaconda. The examples provided with the code allow for a quick start and the Python ecosystem comes with a large user base with lots of support. The Python ecosystem is at the root of the PyPWA coding and parallelism is inherent in its design (either CPU or GPU uses).

\section{Acknowledgements}
We would like to acknowledge former members of the PyPWA project that contributed during the past years: Brandon DeMello, Joshua Pond, Christopher Banks, Michael Harris, Stephanie Bramlett, and Ryan Wright.  We would like to thank the Thomas Jefferson National Facility (JLab) computing division for their support, and Prof. G. Hsieh of Norfolk State University (NSU) Computing Sciences Department, for facilitating the use of their HPC cluster. This work was supported in part by the  National Science Foundation grants \# 0855338, \# 1205763 and \# 2110797, the Science and Technology Facilities Council (STFC) of the United Kingdom, and the U.S. Department of Energy, Office of Science, Office of Nuclear Physics under contract DE-AC05-06OR23177.

{\it Declaration  AI-assisted technologies in the writing process.}\\
During the preparation of this work, the authors used ChatGPT in order to do minor copy editing and suggestions for figure captions. After using this service, the authors reviewed and edited the content as needed and take full responsibility for the content of the publication.






\bibliographystyle{elsarticle-num} 
\bibliography{refs}

\begin{thebibliography}{10}
\expandafter\ifx\csname url\endcsname\relax
  \def\url#1{\texttt{#1}}\fi
\expandafter\ifx\csname urlprefix\endcsname\relax\def\urlprefix{URL }\fi
\expandafter\ifx\csname href\endcsname\relax
  \def\href#1#2{#2} \def\path#1{#1}\fi

\bibitem{Merz}
L.~Schiff, {Quantum Mechanics}, {McGraw Hill}, 1968.

\bibitem{Palano:2022ovo}
A.~Palano, {Light meson spectroscopy and gluonium searches in $\eta_c$ and
  $\Upsilon(1S)$ decays at BaBar}, in: {8th Workshop on Theory, Phenomenology
  and Experiments in Flavour Physics}: {Neutrinos, Flavor Physics and Beyond},
  2022.
\newblock \href {http://arxiv.org/abs/2210.00782} {\path{arXiv:2210.00782}}.

\bibitem{CLAS:2008ycy}
M.~Battaglieri, et~al., {First measurement of direct f0(980) photoproduction on
  the proton}, Phys. Rev. Lett. 102 (2009) 102001.
\newblock \href {http://arxiv.org/abs/0811.1681} {\path{arXiv:0811.1681}},
  \href {https://doi.org/10.1103/PhysRevLett.102.102001}
  {\path{doi:10.1103/PhysRevLett.102.102001}}.

\bibitem{GlueX:2021pcl}
S.~Adhikari, et~al., {Measurement of spin density matrix elements in
  \ensuremath{\Lambda}(1520) photoproduction at 8.2\textendash{}8.8 GeV}, Phys.
  Rev. C 105~(3) (2022) 035201.
\newblock \href {http://arxiv.org/abs/2107.12314} {\path{arXiv:2107.12314}},
  \href {https://doi.org/10.1103/PhysRevC.105.035201}
  {\path{doi:10.1103/PhysRevC.105.035201}}.

\bibitem{Salgado:2013dja}
C.~W. Salgado, D.~P. Weygand, {On the Partial-Wave Analysis of Mesonic
  Resonances Decaying to Multiparticle Final States Produced by Polarized
  Photons}, Phys. Rept. 537 (2014) 1--58.
\newblock \href {http://arxiv.org/abs/1310.7498} {\path{arXiv:1310.7498}},
  \href {https://doi.org/10.1016/j.physrep.2013.11.005}
  {\path{doi:10.1016/j.physrep.2013.11.005}}.

\bibitem{Ketzer:2019wmd}
B.~Ketzer, B.~Grube, D.~Ryabchikov, {Light-Meson Spectroscopy with COMPASS},
  Prog. Part. Nucl. Phys. 113 (2020) 103755.
\newblock \href {http://arxiv.org/abs/1909.06366} {\path{arXiv:1909.06366}},
  \href {https://doi.org/10.1016/j.ppnp.2020.103755}
  {\path{doi:10.1016/j.ppnp.2020.103755}}.

\bibitem{Geant}
S.~Agostinelli, et~al., {GEANT4--a simulation toolkit}, Nucl. Instrum. Meth. A
  506 (2003) 250--303.
\newblock \href {https://doi.org/10.1016/S0168-9002(03)01368-8}
  {\path{doi:10.1016/S0168-9002(03)01368-8}}.

\bibitem{iminuit}
H.~Dembinski, P.~Ongmongkolkul, C.~Deil, H.~Schreiner, M.~Feickert, C.~Burr,
  J.~Watson, F.~Rost, A.~Pearce, L.~Geiger, A.~Abdelmotteleb, A.~Desai, B.~M.
  Wiedemann, C.~Gohlke, J.~Sanders, J.~Drotleff, J.~Eschle, L.~Neste, M.~E.
  Gorelli, M.~Baak, M.~Eliachevitch, O.~Zapata, scikit-hep/iminuit (Mar. 2023).
\newblock \href {https://doi.org/10.5281/zenodo.7750132}
  {\path{doi:10.5281/zenodo.7750132}}.

\bibitem{James:1975dr}
F.~James, M.~Roos, {Minuit: A System for Function Minimization and Analysis of
  the Parameter Errors and Correlations}, Comput. Phys. Commun. 10 (1975)
  343--367.
\newblock \href {https://doi.org/10.1016/0010-4655(75)90039-9}
  {\path{doi:10.1016/0010-4655(75)90039-9}}.

\bibitem{Foreman_Mackey_2013}
D.~Foreman-Mackey, D.~W. Hogg, D.~Lang, J.~Goodman, emcee: The {MCMC} hammer,
  Publications of the Astronomical Society of the Pacific 125~(925) (2013)
  306--312.
\newblock \href {https://doi.org/10.1086/670067} {\path{doi:10.1086/670067}}.

\bibitem{Cummings2003AnOAOriginal}
J.~P. Cummings, D.~P. Weygand, An object-oriented approach to partial wave
  analysis, arXiv:physics/0309052v1 (2003).

\bibitem{compwa}
Compwa, \url{https://compwa-org.readthedocs.io/} (2023).

\bibitem{amptools}
Amptools, \url{https://github.com/mashephe/AmpTools} (2023).

\bibitem{gpupwa}
Gpupwa, \url{https://sourceforge.net/projects/gpupwa/} (2016).

\bibitem{rootpwa}
Rootpwa, \url{https://github.com/ROOTPWA-Maintainers/ROOTPWA} (2020).

\bibitem{Nickolls:2008gqs}
J.~Nickolls, I.~Buck, M.~Garland, K.~Skadron, {Scalable Parallel Programming
  with CUDA}, Queue 6~(2) (2008) 40--53.
\newblock \href {https://doi.org/10.1145/1365490.1365500}
  {\path{doi:10.1145/1365490.1365500}}.

\bibitem{root}
R.~Brun, F.~Rademakers, Root — an object oriented data analysis framework,
  Nuclear Instruments and Methods in Physics Research Section A: Accelerators,
  Spectrometers, Detectors and Associated Equipment 389~(1) (1997) 81--86, new
  Computing Techniques in Physics Research V.
\newblock \href {https://doi.org/https://doi.org/10.1016/S0168-9002(97)00048-X}
  {\path{doi:https://doi.org/10.1016/S0168-9002(97)00048-X}}.

\bibitem{harris2020array}
C.~R. Harris, K.~J. Millman, S.~J. van~der Walt, R.~Gommers, P.~Virtanen,
  D.~Cournapeau, E.~Wieser, J.~Taylor, S.~Berg, N.~J. Smith, R.~Kern, M.~Picus,
  S.~Hoyer, M.~H. van Kerkwijk, M.~Brett, A.~Haldane, J.~F. del R{\'{i}}o,
  M.~Wiebe, P.~Peterson, P.~G{\'{e}}rard-Marchant, K.~Sheppard, T.~Reddy,
  W.~Weckesser, H.~Abbasi, C.~Gohlke, T.~E. Oliphant, Array programming with
  {NumPy}, Nature 585~(7825) (2020) 357--362.
\newblock \href {https://doi.org/10.1038/s41586-020-2649-2}
  {\path{doi:10.1038/s41586-020-2649-2}}.

\bibitem{2020SciPy-NMeth}
P.~Virtanen, R.~Gommers, T.~E. Oliphant, M.~Haberland, T.~Reddy, D.~Cournapeau,
  E.~Burovski, P.~Peterson, W.~Weckesser, J.~Bright, S.~J. {van der Walt},
  M.~Brett, J.~Wilson, K.~J. Millman, N.~Mayorov, A.~R.~J. Nelson, E.~Jones,
  R.~Kern, E.~Larson, C.~J. Carey, {\.I}.~Polat, Y.~Feng, E.~W. Moore,
  J.~{VanderPlas}, D.~Laxalde, J.~Perktold, R.~Cimrman, I.~Henriksen, E.~A.
  Quintero, C.~R. Harris, A.~M. Archibald, A.~H. Ribeiro, F.~Pedregosa, P.~{van
  Mulbregt}, {SciPy 1.0 Contributors}, {{SciPy} 1.0: Fundamental Algorithms for
  Scientific Computing in Python}, Nature Methods 17 (2020) 261--272.
\newblock \href {https://doi.org/10.1038/s41592-019-0686-2}
  {\path{doi:10.1038/s41592-019-0686-2}}.

\bibitem{Hunter:2007}
J.~D. Hunter, {Matplotlib: A 2D Graphics Environment}, Computing in Science \&
  Engineering 9~(3) (2007) 90--95.
\newblock \href {https://doi.org/10.1109/MCSE.2007.55}
  {\path{doi:10.1109/MCSE.2007.55}}.

\bibitem{peterson2009f2py}
P.~Peterson, F2py: A tool for connecting fortran and python programs, Int. J.
  Comput. Sci. Eng. 4~(4) (2009) 296–305.
\newblock \href {https://doi.org/10.1504/IJCSE.2009.029165}
  {\path{doi:10.1504/IJCSE.2009.029165}}.

\bibitem{behnel2011cython}
S.~Behnel, R.~Bradshaw, C.~Citro, L.~Dalcin, D.~S. Seljebotn, K.~Smith, Cython:
  The best of both worlds, Computing in Science \& Engineering 13~(2) (2011)
  31--39.
\newblock \href {https://doi.org/10.1109/MCSE.2010.118}
  {\path{doi:10.1109/MCSE.2010.118}}.

\bibitem{soton403913}
T.~Kluyver, B.~Ragan-Kelley, F.~P{\'e}rez, B.~Granger, M.~Bussonnier,
  J.~Frederic, K.~Kelley, J.~Hamrick, J.~Grout, S.~Corlay, P.~Ivanov, D.~Avila,
  S.~Abdalla, C.~Willing, J.~development team,
  \href{https://eprints.soton.ac.uk/403913/}{Jupyter notebooks - a publishing
  format for reproducible computational workflows}, in: F.~Loizides, B.~Scmidt
  (Eds.), Positioning and Power in Academic Publishing: Players, Agents and
  Agendas, IOS Press, 2016, pp. 87--90.
\newline\urlprefix\url{https://eprints.soton.ac.uk/403913/}

\bibitem{PER-GRA:2007}
F.~P\'erez, B.~E. Granger, \href{https://ipython.org}{{IP}ython: a system for
  interactive scientific computing}, Computing in Science and Engineering 9~(3)
  (2007) 21--29.
\newblock \href {https://doi.org/10.1109/MCSE.2007.53}
  {\path{doi:10.1109/MCSE.2007.53}}.
\newline\urlprefix\url{https://ipython.org}

\bibitem{NEURIPS2019_9015}
A.~Paszke, S.~Gross, F.~Massa, A.~Lerer, J.~Bradbury, G.~Chanan, T.~Killeen,
  Z.~Lin, N.~Gimelshein, L.~Antiga, A.~Desmaison, A.~Kopf, E.~Yang, Z.~DeVito,
  M.~Raison, A.~Tejani, S.~Chilamkurthy, B.~Steiner, L.~Fang, J.~Bai,
  S.~Chintala, Pytorch: An imperative style, high-performance deep learning
  library, in: H.~Wallach, H.~Larochelle, A.~Beygelzimer, F.~d\textquotesingle
  Alch\'{e}-Buc, E.~Fox, R.~Garnett (Eds.), Advances in Neural Information
  Processing Systems 32, Curran Associates, Inc., 2019, pp. 8024--8035.

\bibitem{Cummings2003AnOA}
J.~P. Cummings, D.~P. Weygand, An object-oriented approach to partial wave
  analysis (2003).
\newblock \href {http://arxiv.org/abs/physics/0309052}
  {\path{arXiv:physics/0309052}}.

\bibitem{cryptoeprint:2010/548}
S.~Gueron, S.~Johnson, J.~Walker, Sha-512/256, Cryptology ePrint Archive, Paper
  2010/548, \url{https://eprint.iacr.org/2010/548} (2010).

\bibitem{James:2004xla}
F.~James, M.~Winkler, {MINUIT User's Guide} (2004).

\bibitem{corner}
D.~Foreman-Mackey, corner.py: Scatterplot matrices in python, The Journal of
  Open Source Software 1~(2) (2016) 24.
\newblock \href {https://doi.org/10.21105/joss.00024}
  {\path{doi:10.21105/joss.00024}}.

\bibitem{pep256}
D.~Goodger, \href{https://peps.python.org/pep-0256/}{Docstring processing
  system framework}, PEP 256 (2001).
\newline\urlprefix\url{https://peps.python.org/pep-0256/}

\bibitem{pep287}
D.~Goodger, \href{https://peps.python.org/pep-0287/}{Docstring conventions},
  PEP 287 (2002).
\newline\urlprefix\url{https://peps.python.org/pep-0287/}

\bibitem{docs}
\url{https://pypwa.readthedocs.io/en/main/} (2023).

\bibitem{website}
\url{https://www.jlab.org/pypwa} (2023).

\bibitem{github}
\url{https://github.com/JeffersonLab/PyPWA} (2023).

\bibitem{10.1145/1465482.1465560}
G.~M. Amdahl, Validity of the single processor approach to achieving large
  scale computing capabilities, in: Proceedings of the April 18-20, 1967,
  Spring Joint Computer Conference, AFIPS '67 (Spring), Association for
  Computing Machinery, New York, NY, USA, 1967, p. 483–485.
\newblock \href {https://doi.org/10.1145/1465482.1465560}
  {\path{doi:10.1145/1465482.1465560}}.

\bibitem{Vincent}
V.~Mathieu, M.~Albaladejo, C.~Fern\'andez-Ram\'{\i}rez, A.~W. Jackura,
  M.~Mikhasenko, A.~Pilloni, A.~P. Szczepaniak, Moments of angular distribution
  and beam asymmetries in $\ensuremath{\eta}{\ensuremath{\pi}}^{0}$
  photoproduction at gluex, Phys. Rev. D 100 (2019) 054017.
\newblock \href {https://doi.org/10.1103/PhysRevD.100.054017}
  {\path{doi:10.1103/PhysRevD.100.054017}}.

\end{thebibliography}

\appendix
\section{ \\ Example: $\eta\pi$ photoproduction}\label{sec:example}
To demonstrate the use of PyPWA, we present a partial wave analysis on simulated data of the reaction $\overrightarrow{\gamma} p \rightarrow  \eta \pi \ p  $, the photoproduction of two pseudo-scalar mesons. The goal is to find intermediary meson states (X) that decay via $X \rightarrow \eta \pi$. We assume that the reaction 
$\overrightarrow{\gamma} p \rightarrow X p $ occurs by diffractive scattering of a linearly polarized photon beam off a proton target which remains intact. All code necessary to run this example is available on our GitHub page~\cite{github}.

We perform a so-called \textit{mass-independent} partial wave analysis~\cite{Salgado:2013dja}, where we include the four-momentum transfer dependency of the production amplitudes in our kinematic simulation. We divide the data and simulation in bins of $X$ meson mass and consider fixed beam energy. Therefore, the two pseudo-scalar intensities, in each mass bin, will have only angular dependencies.
We factorize the total transition amplitudes into a production amplitude $V$ (interaction, $X$ production) and a decay amplitude $A$ (decay of $X$ into the final state particles). Production amplitudes are generally unknown at these energies, they are fully (mass-independent) or partially (mass-dependent) fitted to the data.  The production amplitudes contain information about the hadronic QCD-based interaction. Those interactions are more difficult to model; therefore, for a mass-independent analysis, the production amplitudes will be considered a {\it weight}  on each decay amplitude.  These weights are the parameters to be fitted to the data to obtain the observed overall intensity. Classic partial wave decomposition, truncated to low angular momenta contributions, represents a good first attempt to obtain a set of  amplitudes, {\it the model}.  It can then be checked that the data are reasonably described by this model. It can be shown \cite{Salgado:2013dja}, that for two pseudo-scalars, the decay amplitudes are given by spherical harmonics.

\begin{table}[]
\centering
\caption{Resonances/Waves definition for simulation, See text for details.}
\label{tab:Table_I}
\begin{tabular}{ccccc}
\hline
mass (GeV/$c^2$)     & weight &         ($\epsilon,L,M$)  
             \\ \hline
0.980 & 0.5 & (1,0,0)  \\
1.306 & 0.3 & (1,2,1) \\
1.722 & 0.2 & (1,1,1) \\
\hline
\end{tabular}
\end{table}
\begin{figure}[!htp]
    \centerline{
    \includegraphics[width=\linewidth,trim={.5cm .8cm .8cm 0.3cm},clip]{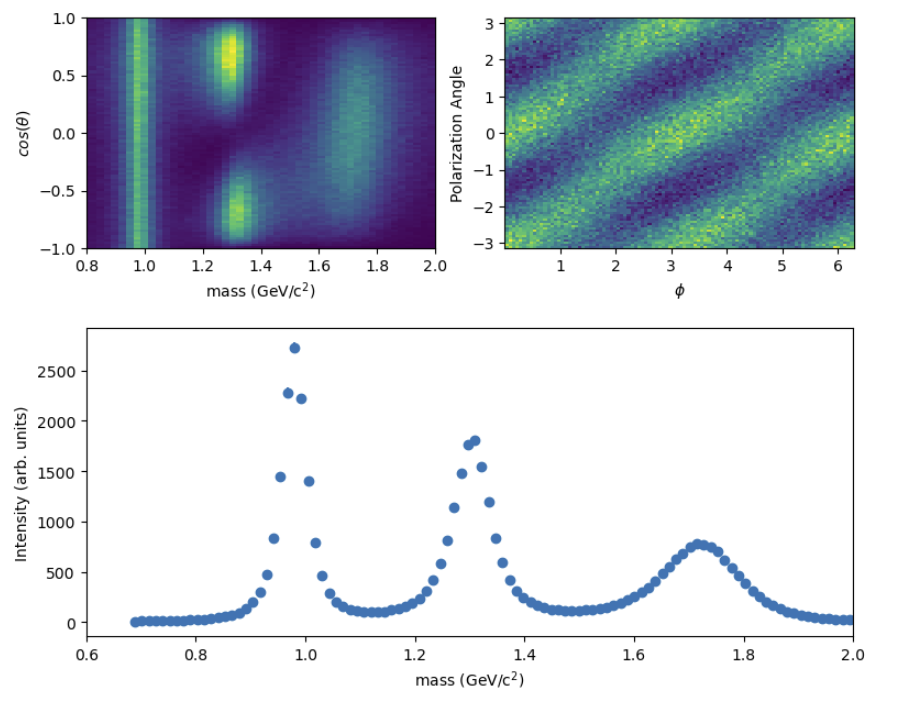}}
    \caption{Generated mass distribution showing the three input resonances. Also shown are the  $\cos(\theta)$ vs. $mass$ and $polarization$\ $angle $ vs. $\phi$ for the same generated data. Angles $(\theta,\phi)$ are defined in the Helicity frame.}
    \label{fig:gen}
\end{figure}

 We use amplitudes derived by the JPAC \cite{Vincent} collaboration to simulate and consecutively fit the intensities of the given set of resonances and waves. We use a beam with a fixed energy of 8 GeV and 40\% linear polarization fraction and an exponential t-Mandelstam distribution $\propto\exp\left(-bt\right)$ with $b=\SI{6}{\GeV^{-2}}$. 
The JPAC two pseudo-scalar amplitudes are defined by the following parameters: $\mathscr{P}$ is the polarization fraction, $\Phi$ is the polarization angle, and $\epsilon, L,m$ are the three quantum numbers of the waves (reflectivity, angular momentum and positive z-component of angular momentum). The two angles $\theta , \phi$ are  describing the decay particles in the Helicity frame \cite{Vincent}, the total intensity is given by Equation~\eqref{eq:apploss} (Equation~(B4) of reference~\cite{Vincent}).

\begin{equation*}
  I(\theta, \phi,\mathscr{P},\Phi) = 
I^{(0)}(\theta, \phi) -\mathscr{P}I^{(1)}(\theta, \phi)\cos 2\Phi
\end{equation*}
\begin{equation}
-\mathscr{P}I^{(2)}(\theta, \phi)\sin 2\Phi
\label{eq:apploss}\end{equation}
The intensities, $I^{(0)},I^{(1)},I^{(2)}$ are calculated from expressions (D12) of Appendix D of reference~\cite{Vincent}.

An example of code used to calculate the JPAC intensities (using PyTorch) is shown below (see full code on GitHub~\cite{github}):
\begin{lstlisting}
# Calculate Total Intensity
import numpy as np
import pandas as pd
import scipy.special
import torch as tc
from PyPWA import NestedFunction

# Calculate Decays directly from Spherical Harmonics
def produce_specific_decay(theta, phi, m_waves, l_waves):
    theta[theta < 0] = theta[theta < 0] + 2 * np.pi

    decay = np.empty((len(m_waves), len(theta)), "c16")
    for index, (m, l) in enumerate(zip(m_waves, l_waves)):
        decay[index] = scipy.special.sph_harm(m, l, theta, phi)

    return decay

class FitWithGPU(NestedFunction):

    USE_TORCH = True
    USE_MP = False

    def __init__(self, initial_params):
        super(FitWithGPU, self).__init__()
        self.device = tc.device('cpu')
        self.__alpha = tc.Tensor([])
        self.__pol = tc.Tensor([])
        self.__phi = np.array([])
        self.__theta = np.array([])
        self.__decay = tc.Tensor([])

        elm = make_elm(initial_params)
        self.__e = tc.from_numpy(elm['e'])
        self.__l = elm['l']
        self.__m = elm['m']

    def setup(self, data):
        # Handle Torch Devices based on USE_MP status
        if not self.USE_MP:
            self.device = tc.device("cuda:0")
        self.__alpha = tc.from_numpy(data['alpha']).to(self.device)
        self.__pol = tc.from_numpy(data['pol']).to(self.device)
        self.__e = self.__e.to(self.device)

        self.__decay = tc.from_numpy(produce_specific_decay(
            data['phi'], data['theta'], self.__m, self.__l
        )).to(self.device)
        return self

    def calculate(self, params):
        vs = params[::2] + 1j * params[1::2]
        vs = tc.from_numpy(vs).to(self.device)

        v = vs * self.__decay.T
        v_conj = vs * tc.conj(self.__decay).T

        u10 = v.T[self.__e == 1].sum(0)
        u20 = v_conj.T[self.__e == 1].sum(0)
        u11 = v.T[self.__e == -1].sum(0)
        u21 = v_conj.T[self.__e == -1].sum(0)

        return self.__compute_intensity(u10, u20, u11, u21)

    def __compute_intensity(
            self, u10: tc.Tensor, u20: tc.Tensor,
            u11: tc.Tensor, u21: tc.Tensor) -> tc.Tensor:

        i0 = (
                u10 * tc.conj(u10) + u20 * tc.conj(u20)
                + u11 * tc.conj(u11) + u21 * tc.conj(u21)
        ).real

        i1 = 2 * (-1 * (u10 * tc.conj(u20)).real 
            + (u11 * tc.conj(u21)).real)
        i2 = 2 * (-1 * (u10 * tc.conj(u20)).imag 
            + (u11 * tc.conj(u21)).imag)

        intensity = i0 - self.__pol * i1 * tc.cos(2 * self.__alpha)
        intensity -= self.__pol * i2 * tc.sin(2 * self.__alpha)

        return intensity.real
\end{lstlisting}
We perform the analysis on a simulated sample and assume a detector with full acceptance and 100\% efficiency. Using our simulation package, we simulated three resonances decaying to $ \eta \pi$, each in a different pure wave, defined by the reflectivity ($\epsilon$), total angular momentum ($L$) and z-component of the angular momentum ($m$). The input values are listed in Table~\ref{tab:Table_I}.

The simulated $X$ mass distribution, the cosine of the polar helicity angle of the $\eta$ particle ($\theta$) versus $X$ mass, and the polarization angle versus the azimuthal helicity angle ($\phi$) of the $\eta$ particle are shown in Figure~\ref {fig:gen}. In this example, our analysis goal is to extract the resonances from the simulated data by fitting the $\eta \pi$ angular distributions using the same JPAC amplitudes.

\begin{figure}[!htp] 
    \centerline{
    \includegraphics[width=\linewidth,trim={0.1cm .1cm 0.8cm 0.8cm},clip]{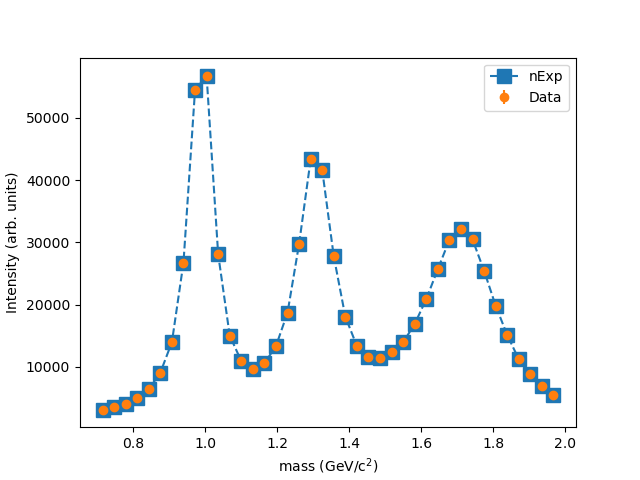}}
    \caption{The figure shows the fitted total Intensity (squares) versus mass. The simulated data points are also plotted (circles). The fitted values and data are in very good agreement}
    \label{fig:fitmass1}
\end{figure}

We define the set of waves we will use in the fit and initial values for the production amplitudes. The optimization will be done via an event-by-event extended log-likelihood fit using iMinuit.
In this example, we used only positive reflectivity ($\epsilon=1$), $L<3$ and $M=0,1\ or\ 2$, a S, P, and D wave set. We also need to use simulated (accepted and generated) samples to obtain the expected and true number of entries in a bin to calculate the extended log-likelihood. 
The software package provides tools for the user to define the bins, the variable to be binned, the bin ranges, and the extended log-likelihood.

\begin{figure}[!htp] 
    \centerline{
    \includegraphics[width=\linewidth,trim={0.1cm .1cm 0.8cm 0.8cm},clip]{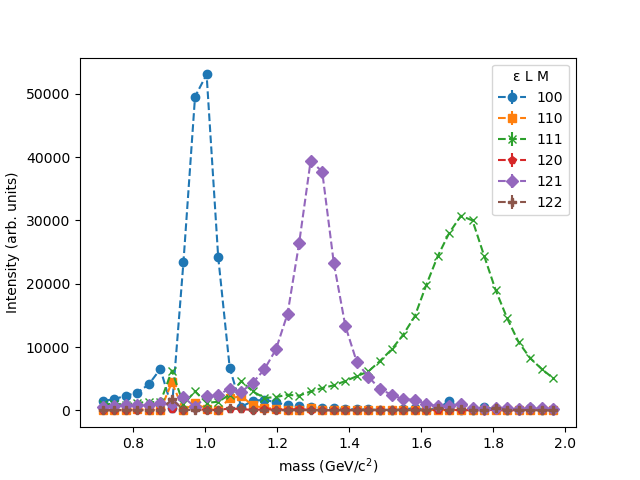}}
    \caption{The figure shows the fitted intensity versus mass for different waves in the fit set. The fit reproduces the wave composition of the simulated data. Added waves, not present in the simulated data, do not contribute to the intensity.}
    \label{fig:fitmass2}
\end{figure}

Fitting with iMinuit, we obtain the optimal production amplitudes that minimize the negative extended log-likelihood (c.f. Equation~\eqref{eqn:like}). 

Calculating the expected number of events in a mass bin, we can compare with the simulated data as shown in Figure~\ref{fig:fitmass1}. 
The fitted values and data are in very good agreement. 
Figure ~\ref{fig:fitmass2} shows the intensities versus mass separated in each of the fitted waves, the goal of our analysis. As shown in Figure~\ref{fig:fitmass2}, the results of the fit match the input resonances and waves of the simulated data well, i.e, we were able to extract resonances and associated quantum numbers (waves) from the simulated input data.

\newpage
\section{\\ Benchmark System Specifications} \label{sec:specs}
Table~\ref{tab:specs} shows the specifications of the Dual Cascade Lake Server used for benchmarking the scaling of PyPWA.
\begin{table}[ht]
\centering
\caption{Specifications of the Dual Cascade Lake Server}
\label{tab:specs}
\begin{tabular}{p{4cm}p{4cm}}
\hline
\multicolumn{2}{c}{\textbf{CPU}} \\
\hline
Model & Intel Xeon Gold 6230 \\
Architecture & Cascade Lake-SP \\
Socket & Dual Socket LGA-3647 \\
Cores/Threads & 20 / 40 per CPU \\
Base Frequency & 2.1 GHz \\
Max Turbo Frequency & 3.9 GHz \\
Cache & 27.5 MB\\
\hline
\multicolumn{2}{c}{\textbf{Memory}} \\
\hline
Type & DDR4-2933 ECC \\
Capacity & 512 GB (16 x 32GB) \\
\hline
\multicolumn{2}{c}{\textbf{GPU}} \\
\hline
Model & NVIDIA Tesla V100 PCIe \\
Architecture & Volta \\
Number of GPUs & 3 \\
CUDA Cores & 5120\\
Memory & 32 GB HBM2\\
\hline
\end{tabular}
\end{table}

\end{document}